# Large area buried nanopatterning by broad ion implantation without any mask or direct writing

Prasanta Karmakar[1*], Biswarup Satpati[2]

[1]Variable Energy Cyclotron Center, 1/AF, Bidhannagar, Kolkata -700064, India
[2]Saha Institute of Nuclear Physics, 1/AF, Bidhannagar, Kolkata -700064, India

We have introduced here a simple, single step and cost effective broad ion beam technique for preparation of nanoscale electronic, magnetic, optical and mechanical devices without the need of resist, mask, or focused electron and ion beams. In this approach, broad beam ion implantation of desired atom on a prefabricated ion beam patterned surface promotes site selective deposition by adjusting the local angle of ion implantation. We show that implantation of Fe ions on an $O^+$ induced pre fabricated triangular shaped patterned Si surface results in a self-organized periodic array of striped magnetic nanostructures having several micron length and about 50 nm width arranged with a lateral separation of ~ 200 nm. The morphology, composition, crystalline structure and magnetic property of these nanopatterns have been analyzed and the crucial feature of such nanopattern formation by broad beam ion implantation is explained.

PACS number(s): 61.80.Jh, 81.16.Nd, 79.20.Rf, 68.37.Og,

*Corresponding author e-mail: prasantak@vecc.gov.in.

Development of a system with nanometer scale periodic array is technologically important because of its unique magnetic, electronic, optical as well as mechanical properties. The development of such system relies on the technique that can manipulate with atomic precision. Therefore modern technique like AFM/STM tip or dip pen lithography is used to realize such structure. However, these approaches of nanopatterning are very slow and expensive for large area and mass production. Alternative to this is to use nanometer focused beam of ions, electrons or photon which can produce structures order of magnitude faster than mechanical tips[1]. However, patterning with focused beam is not cost and time effective for large scale fabrication. Photolithography is conventionally used for large scale patterning but the spatial resolution of patterning in this case is restricted by optical diffraction limit. X-ray lithography is promising compared to photo lithography for nano scale patterning, but it is expensive and requires huge x-ray sources as well as resist, mask and also multi step processing. Similarly, broad beams of ions and electrons could also fabricate nanopatterns using resist, mask and multi step processing [2, 3]. Therefore, search for a new technique for fast and large area nanopattern formation in minimum steps without mask and lithography remains an important issue in materials science research.

Broad ion beam induced nanopatterning is believed to be one of the potential candidates for large scale low cost nano fabrication but nano patterns formed by this technique has some limitation. The patterned surface structure is an integral part of the bulk material and is difficult to isolate from it. For isolation, patterns could be formed on thin films, supported on substrate [4], however, the substrate material may mix with the structure during sputtering at the interface. Alternatively, prefabricated large scale nanopatterns could be used as template to transfer the pattern on other material by deposition, but isolated stripe or dot formation is difficult by deposition only, because the atoms having thermal energy are deposited in an uncontrolled way on the patterned surface[5, 6]. Ion beam implantation on prefabricated nanopatterns could overcome these short falls because of its directionality and precise control over depth.

Here, we have developed a new method that allows patterning of isolated nano arrays by broad ion implantation on ion beam induced prefabricated ripple patterned substrate without any mask and/or lithography. Large area patterned Si substrate with sinusoidal or triangular surface profile have been first fabricated by broad Oxygen or Carbon ion beam bombardment at $60^o$ angle of incidence with respect to the surface normal. We show that further oxygen, Carbon or Fe implantation at same ion incidence angle on such pre-formed patterned substrate could selectively deposit projectile atoms at the steep-edged slope (facing the ion beam) of the sinusoidal or triangular ripples. Consequently, one gets a self-organized periodic array of conducting and magnetic striped nanostructures having several micron lengths and a typical width of ~ 50 nm arranged with a lateral separation of ~ 200 nm. The crucial feature of this method is that the projectile atoms are restricted to insert and retain only at specific sites without requiring any mask. Exploiting the geometrical shape of these structures and resulting local ion impact angle variation, site specific implantation of suitable ions, conductive and magnetic patterns were produced on entire beam exposed area (1cm × 1cm in these cases). We have performed detail analyses of these magnetic and conducting nanostructures with respect to their morphology, composition, crystalline structure and magnetic property using cross sectional Transmission Electron Microscopy (XTEM) and associated techniques like energy dispersive X-ray (EDX) spectroscopy, high-angle annular dark field scanning/ transmission electron microscopy (STEM-HAADF) energy-filtered TEM (EFTEM) and Magnetic Force Microscopy (MFM).

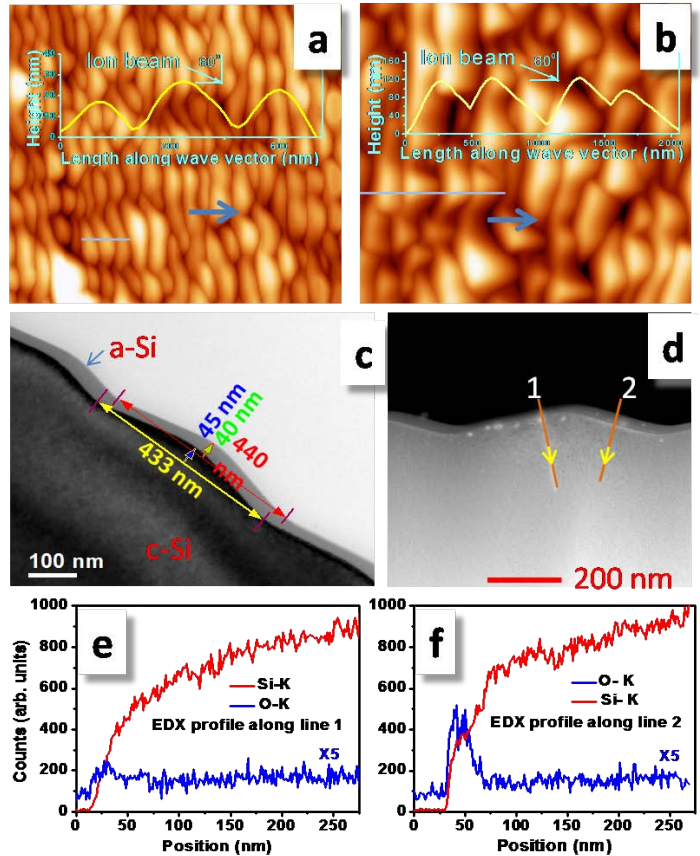

**Fig. 1**. **Atomic force and cross-sectional transmission electron microscopy of 8 keV $O^+$ ion induced patterned Si surfaces**. (**a** to b) AFM images and line profiles along the marked lines, arrows indicate the projection of ion beam direction. (**a**) Sinusoidal ripple structure. (**b**)Triangular faceted structure. (**c** to **d**) cross sectional TEM view of sinusoidal ripples formed at ion fluence of $7\times10^{17}$ ions/cm$^{-2}$. (**c**) Low magnification image. (**d**) STEM-HAADF image. (**e** to **f**) EDX line profiles for Si and O distribution across the structure. (**e**) EDX profile along for line 1. (**f**) EDX profile along for line 2.

Cleaned and degreased Si(100) substrates were bombarded with broad (10 mm) 10 keV mass analyzed Oxygen or Carbon beams at oblique angle ($60^0$) with respect to the surface normal to form periodic ripple structures on the entire ion exposed area. To create spatially resolved oxide or carbide structure further oxygen or carbon ion beams were implanted at same oblique ion incidence angle (Fig 1a). Iron stripes were produced by subsequent 36 keV $Fe^{3+}$ ions implantation on the spatially resolved oxidized face of the ripple at same ion incidence angle (Fig 1b). The mass analyzed beams were extracted from 6.4 GHz ECR based ion beam system[7] and also using Low Energy Ion Beam facility[8]. Cross-sectional TEM (XTEM) specimens were prepared using the standard method of mechanical grinding and double dimpling with final thinning using a precision-ion- polishing system (PIPS, Gatan, Pleasanton, CA). The ion polish was carried out at 3.0 keV energy without liquid nitrogen cools and followed by a 1.2 keV cleaning process. For TEM observation, specimens were aligned on [110] zone axes TEM investigations were carried out using a FEI, Tecnai $G^2$ F30, S-Twin microscope operating at 300 kV. STEM-HAADF is employed here using the same microscope, which is also equipped with a scanning unit and a HAADF detector from Fischione (model 3000). Energy filtered images were acquired using Gatan Imaging Filter (model 963). The compositional analysis was performed by energy dispersive X-ray spectroscopy (EDS, EDAX Instruments) attachment on the Tecnai $G^2$ F30. Bruker made multimode-V Atomic Force Microscopy (AFM) and Magnetic Force Microscopy (MFM) was used for topographic and magnetic measurements.

Broad ion beam bombardment leads to the formation of dot, ripple, sawtooth and tri angular wave like structure on semiconductor, metal and insulator surfaces [9, 10]. Fig. 1a and 1b show the ripple and triangular structure formation by 8 keV ion bombardments on Si (100) surface at oblique angle. Our previous studies of ripple formation by $O^+$ ion bombardment at oblique incidence show that regular ripple pattern on Si starts to form at an ion fluence of $\sim 5\times10^{17}$ ions/ cm$^2$ on entire ion exposed area [11]. Continued bombardment of oxygen ion on the same surface at the same geometry results preferential implantation of oxygen on the front face of the ripples and gets more oxidized compared to the back face. Conducting AFM measurements showed the periodic conducting- semiconducting arrays on the surface [12].

Fig. 1c and 1d show a representative XTEM image and STEM-HAADF image of the prefabricated ripple structures generated by oxygen ion bombardment on Si at oblique angel ($60^0$). Fig. 2e and 2f represents the EDX line profiles along the line 1 and 2, respectively drawn in Fig. 2d. One can see clearly from these cross-sectional images and EDX line profile that oxygen is preferentially implanted in the front side (side of line 2) and forms the non-conducting zones only on the front side keeping the back side semiconducting. EDX line profile also shows that the almost double quantity of oxygen ions were implanted in the beam facing side compared to the rear side. This result suggests the exciting possibilities of changing conducting properties locally of a sample in large scale without any mask and lithography.

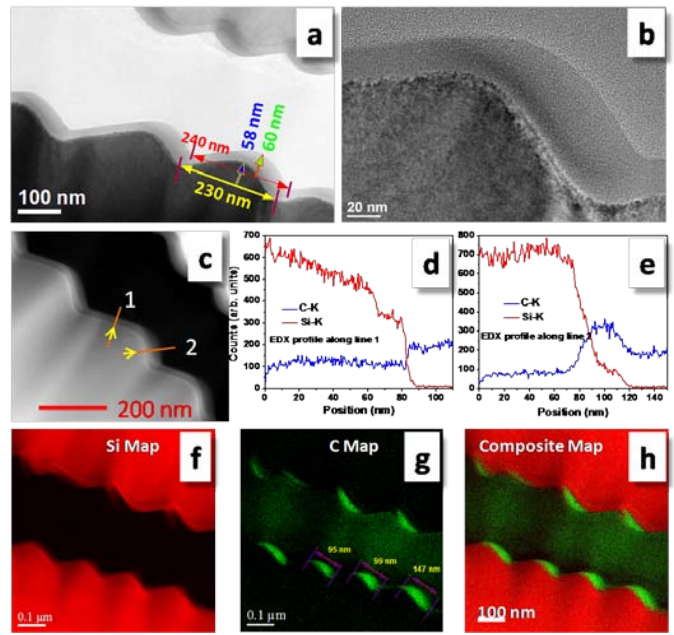

**Fig. 2. Cross-sectional Transmission electron microscopy of 8 keV $C^+$ ion induced patterned Si surface at a fluence of $7\times10^{17}$ ions $cm^{-2}$.** (**a to c**) TEM images. (**a**) Low-mag TEM image. (**b**) HRTEM image. (**c**) STEM-HAADF image. (**d to e**) EDX line profiles for Si and C distribution across the structure. (**d**) From back side along line 1. (**e**) Front side along line 2. (**f to h**) Cross-sectional EFTEM images taken from another region. (**f**) Chemical map of Si (red). (**g**) Chemical map of C (green), and (**h**) composite image showing Si (red) and C (green), indicating the locations of C atoms at the ripple surface.

A similar system was generated by carbon ion beam bombardment on Si (100) surface. Carbon ion bombardment generates the large area ripple structures on Si by oblique angle bombardment at a fluence of ~ $5\times10^{17}$ ions/cm$^2$ [13]. Further bombardment of carbon ion on the preformed ripple leads to preferential implantation on the front side of the ripple. This produces repeated adjoining regions of silicon-carbide and silicon. Fig. 2 shows the cross sectional views of the ripples where carbon rich front regions are visible from XTEM image and EDX profile.

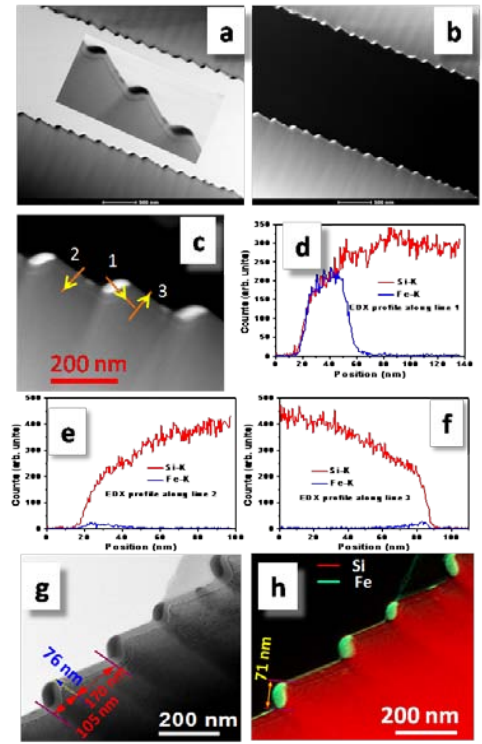

**Fig. 3. Cross-sectional Transmission electron microscopy of the patterned Si surface produced by 8 keV O$^{1+}$ ions with fluence $2\times10^{18}$ ions cm$^{-2}$ and consequent 36 keV Fe$^{3+}$ implantation with fluence $1\times10^{18}$ ions cm$^{-2}$.** (**a** to **b**) STEM images. (**a**) Bright-field image and high-magnification image in the inset. (**b**) Corresponding HAADF image. (**c**) High-magnification STEM-HAADF image. (**d** to **f**) EDX line profiles for Si and Fe distribution across the structure. (**d**) EDX profile for line 1 in C. (**e**) EDX profile for line 2 in C. (**f**) EDX profile for line 3 in C. (**g** to **h**) EFTEM images. (**g**) Zero-loss image. (**e**) Composite image showing Si (red) and Fe (green).

The above results stimulate us to generate magnetic array by implanting Fe ion on pre fabricated ion beam sputtered faceted ripple morphology of Si. The ripples were first formed by 8 keV oxygen ion bombardments as displayed in Fig. 1. Additional bombardment (fluence) of oxygen ions changed the shape of the sinusoidal ripple profile into triangular wave like structure (Fig 1b). The transformation of an initial sinusoidal shape of IBS sputtered ripple into a triangular like profile after prolonged bombardment (high fluence/dose) was also observed before [11, 14, 15]. Due to high dose of oxygen ion implantation at the steep-edged faces of the triangular structure, an amorphous silicon oxide layer of thickness 64 nm was formed as shown in Fig. 4. A similar observation has also been reported for a high dose Ar ion beam sputtered Si surface [16]. Now, on the prefabricated oxygen sputtered Si surface having triangular shaped profile, Fe ion beams were implanted at an angle $60^0$ with respect to the global surface normal. The incident beams are only implanted on the sharp-rise face of the triangular ripple structure where prior oxygen ion bombarded amorphous oxide layer was present. Thus large area isolated periodic stripe of magnetic structures on silicon oxide matrix was generated. Fe was rarely implanted on the back surface of the ripples because of the grazing incidence ion bombardment on back surface. XTEM image shows the side view of the implanted iron in the silicon oxide matrix (Fig. 3). Almost no Fe is implanted on the back surface whereas pony bean like structures are visible in the front side. EDX line profile on cross sectioned sample clearly shows that the Fe only sits on the front surface whereas Fe signal is almost negligible on the back side (Fig. 3.

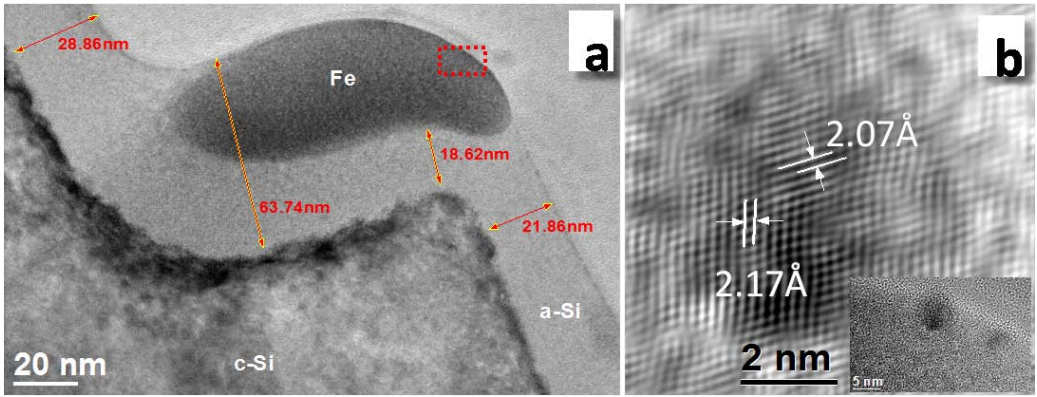

**Fig. 4. High resolution cross-sectional TEM images.** (**a**) Pony bean like Fe particle seating on the front side patterned Si surface. (**b**) Fourier filtered image from edge of Fe particle (shown in inset) from dotted box region in A showing lattice fringes and spacing close to (100) and (002) inter planer spacing of hexagonal Fe [JCPDS # 65-5099].

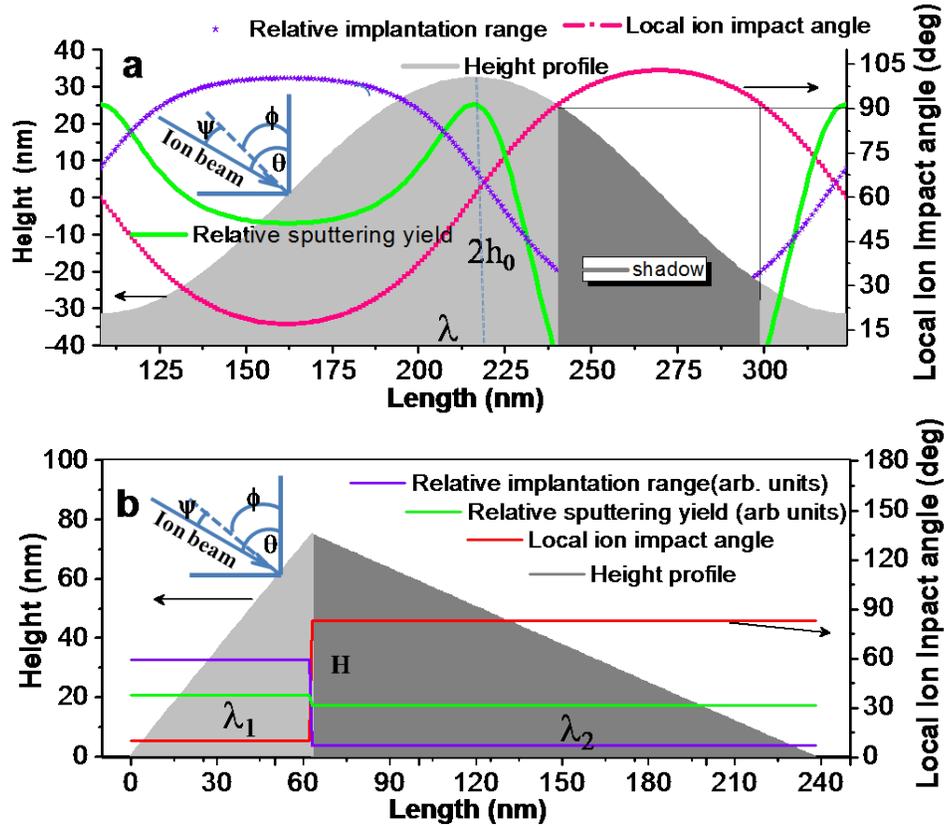

c-f).

**Fig. 5. Variation of local ion impact angle, ion implantation range and sputtering yield with length along the wave vector direction.** (**a**) A sinusoidal ripple with amplitude $h_0 = 32$ nm and wavelength $\lambda = 275$ nm similar to the structure shown in Fig. 1 & 2. (**b**) A triangular faceted structure with height $H = 76$ nm with $\lambda_1 = 105$ nm and $\lambda_2 = 170$ nm similar to the structure shown in Fig. 3. The ion incidence angel with respect to the global normal is $\theta = 60^0$ and $\psi$ is ion impact angle with respect to the local normal.

The spatial variation of implanted species over the sinusoidal ripple structure and triangular wave like structure has been described by calculating the local ion impact angle, implantation depth and sputtering yield. It is found that a certain portion of the structures are not exposed by the ion beam. As a first approximation, we can consider that initial stage of the ion bombardment forms the sinusoidal ripple structure with the surface profile described by $h = h_0 \cos\left(\frac{2\pi x}{\lambda}\right)$ where $h$ is the height, $x$ is distance along the ripple wave vector, $h_0$ is the amplitude and $\lambda$ is the wavelength of the ripple. In the present case, with respect to the Fig. 1c, the peak to peak height is defines as $2h_0 = 40$ nm, meaning the amplitude of the ripple here is 20 nm for the ripple structure depicted. The slope at any arbitrary point is $\frac{dh}{dx} = \frac{2\pi h_0}{\lambda}\sin\left(\frac{2\pi x}{\lambda}\right)$. If we consider the ion incidence angle is $\theta$ with respect to the global normal of the surface and $\phi$ is the angle between global normal and local normal then, $\tan\phi = \frac{2\pi h_0}{\lambda}\sin\left(\frac{2\pi x}{\lambda}\right)$. Therefore, the ion incidence angle with respect to the local surface normal could be written for the above sinusoidal ripple structure as $\psi = \theta - \tan^{-1}\left[\frac{-2\pi h_0}{\lambda}\sin\left(\frac{2\pi x}{\lambda}\right)\right]$ …..(1). The variation of ion incidence angle at different point of the ripple with respect to the local normal is presented in Fig 5. It is found that depending on the ratio of amplitude and wavelength (aspect ratio) of the sinusoidal ripple structures, the ion incidence angle changes along the direction of ripple wave vector. In the first case (Fig. 1c) when average ripple amplitude is 20 nm and wavelength is 440 the ion incidence angle with respect to local surface normal varies from $44^0$ to $77^0$, however entire ripple is exposed by ion beam. The aspect ratio (amplitude: wavelength) changes with further ion bombardment and when the aspect ratio is greater than 0.091 in the present case, a certain portion of the sample is not at all exposed by ion beam which we called zone of no radiation. The zone is depicted in Fig. 5a. From equation (1), condition for no beam exposure (shadowing) is $\psi \geq \pi/2$ i.e. $\tan(\pi/2 - \theta) \leq \frac{2\pi h_0}{\lambda}\sin\left(\frac{2\pi x}{\lambda}\right)$, Similar shadowing condition was also predicted by Carter et al.[17].

The case is illustrated in Fig 5b when sinusoidal ripples are transformed into tri angular facets, and Fe ions are implanted at same ion incidence angle ($60^0$) with respect to global normal.

Fig. 5b shows the local ion impact angle on two faces of the tri angular wave. In the front face the local ion impact angle is $\psi_{front} = \theta - \tan(H/\lambda_1)$, whereas at the back face, it is $\psi_{rear} = \theta + (\tan H/\lambda_2)$, where $H$ is the height, $\lambda_1$ and $\lambda_2$ are the bases of two triangles as shown in figure (Fig 5b).

Depending upon the ion impact angle, the penetration depth, collision cascade shape and sputtering yield changes. The penetration depth can be written as $R = \left(\frac{\ln E_i - 1}{c}\right) \cos \theta$, where $E_i$ is the ion energy and c is constant. This equation shows that the depth of an incident ion under off-normal bombardment is equal to the cosine of the incident angle times the depth of the incident ion with the same energy under normal incidence [18]. The implantation depth variation along the ripple wave vector direction is shown (Fig 5 a & b). It is observed from the figure that projectile is penetrated and implanted at larger depth at the front side than that of rear side of the ripple or triangular structure. If the aspect ratio is sufficient to shadow the beam, no implantation takes place at a certain portion on the back side.

The sputtering yield also changes with ion impact angle. Sigmund et al [19] showed the dependence of sputtering yield with ion incidence angle as $\frac{Y(\theta)}{Y(\theta=0)} = (\cos \theta)^{-f_s}$ where the exponent $f_s \approx 1{\sim}2$ depending on the mass of incident ion and target atom. Here the sputtering yield increases with incidence angle and goes to infinity for grazing incidence. More practical expression $\frac{Y(\theta)}{Y(\theta=0)} = (\cos \theta) \exp\left(\frac{a^2 \sin^2 \theta}{2\alpha^2}\right)$ where a and α is the penetration depth and longitudinal straggling respectively, as given by Wei et al[18]. The sputtering yield variation along the wave vector direction is shown in Fig. 5a and 5b for sinusoidal and triangular wave structures, respectively. It is found that the dynamic balance between implantation and sputtering of both sinusoidal and triangular wave structures leads to preferential retention of the projectile where local ion impact angle is normal and near normal. As a result the front side of the structures is only enriched with the projectile.

The estimation of implant enriched zone of the prefabricated nano structures based on the geometrical model (Fig 5) is compared with the observed cross sectional TEM and EDX data. If wavelength and amplitude of the carbon induced ripple as shown in Fig. 2a is 240 nm and 30 nm respectively, the expected length of the shadow ($\psi > 90^0$) and grazing incidence ($70^0 < \psi < 90^0$) zone projected on ripple wave vector direction would be 60 nm and 40 nm, respectively. Therefore, the expected projected length of the implantation zone where implantation range is maximum and sputtering is minimum would be 96 nm (Fig 5a). The observed carbon enriched zone as shown in Fig. 2g is ~ 99 nm which is consistent with the calculated value. In case of tri angular faceted structure the average wavelength is 275 nm and height is 76 nm where $\lambda_1 = 105$ nm and $\lambda_2 = 170$ nm similar to Fig. 3g. The calculation as shown in Fig. 5b yields the front face and rear face ion impact angles $24^0$ and $84^0$ respectively. Low impact angle at the front side of the ripple causes higher penetration depth and low sputtering angle. Consequently, most of the Fe ions would sit only at the front slope of ripple as shown in the cross sectional TEM (Fig. 3). On the otherhand, due to the almost grazing impact, almost no Fe is seen at the rear side.

By co-deposition of Fe during Ar ion bombardment on Si Cubero et al [4] reported the formation of discontinuous Fe rich silicide regions on both sides of the faceted ripple structure. However, if broad beam ion implantation is implemented on a prefabricated ion beam sputtered ripple patterned surface where the angle of ion incidence is selected by exploiting the proper geometrical shape to enhance the local ion penetration depth one can achieve site selective metal deposition, a remarkable observation of the present study and such a method of synthesis of nanostructure has not been reported in the literature to the best of our knowledge. It is also to be mentioned here that as Fe is implanted at the sub-surface layers avoiding the possibility of being contaminated through air exposure, there is no need to put cap layer to protect the pattern.

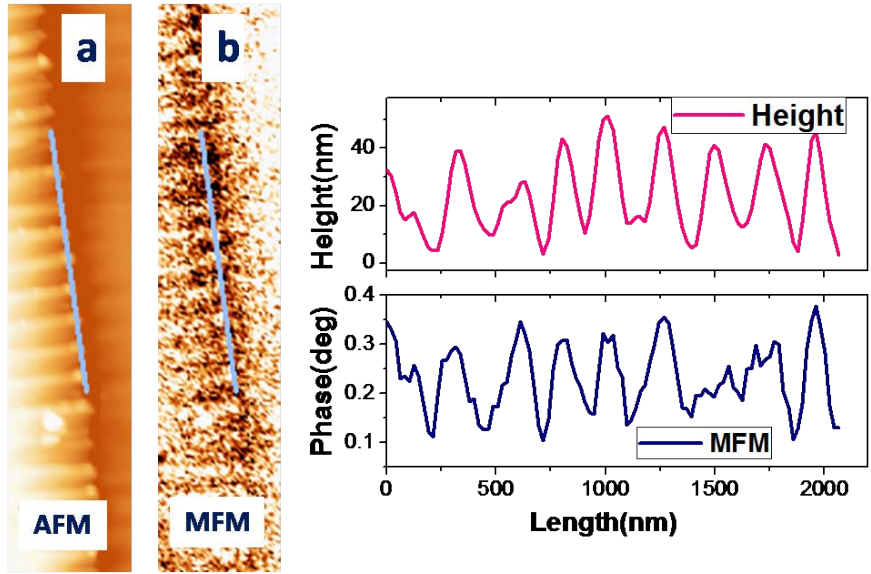

**Fig.6. Cross-sectional Atomic Force and Magnetic Force Microscopy of the 36 keV Fe implanted patterned Si surface.** Topographic image. (**b**) Magnetic force image. Corresponding topographic (upper) and magnetic (lower) profile along the line drawn on (**a**) and (**b**), respectively.

MFM measurements of the Fe implanted magnetic patterns are shown in Fig. 6. Cross-sectional topographic and MFM measurements were performed on the same sample prepared for cross sectional TEM. The topographic images are consistent with the TEM images. The iron implanted zones are visible with the MFM measurement. The magnetic signal is obtained because the implanted Fe in the SiOx matrix rarely formed the silicide. High resolution TEM lattice image shown in Fig. 6, also shows that the implanted Fe clusters are crystalline in nature. Hayashia et al. [20] reported the formation of BCC α-Fe(110) by grazing angle X-ray diffraction of kev energy Fe implantation in $SiO_2$. When the size is about 16 nm super paramagnetic behavior was reported but ferromagnetism appears with increasing Fe dose [20]. Here the height of the stripe is typically 61 nm and elongated in micron size. Therefore, ferromagnetic nature of implanted Fe clusters is expected and MFM could measure the magnetic structures as shown in Fig 6.

The above results demonstrate the potential for engineering nanoscale electronic, magnetic, optical and mechanical devices. In addition, this method not only can be applied for the production of large area stripe pattern but also using the shadowing effect of prefabricated dot pattern isolated magnetic dot or nanostructure of desirable conductivity can be realized. Furthermore, this process of large area nanopatterning with different physical and chemical properties can be realized very easily as the ion beam technique is highly flexible and controllable in nature with respect to ion species, energy and the amount of incorporated material of interest and thus could be a potential technique for cost effective mass fabrication. p. Further, the present method of synthesis lies in the fact that as the desired atomic species is implanted at the sub-surface layers avoiding the possibility of being contaminated through air exposure, there is no need to put cap layer to protect the fabricated nanopatterned structure.

The Authors thank Mrs Sampa Bhattacharjee for her assistance during ion implantation and Prof. T. K. Chini for critical reading of the manuscript and fruitful suggestions. The authors would like to acknowledge Dr. Alok Chakraborti and Dr Vaishali Naik for their constant support.